\documentclass[aps,pra,twocolumn,superscriptaddress,showpacs]{revtex4-1}

\usepackage[pdftex]{graphicx} 
\usepackage{amsmath} 
\usepackage{amssymb} 
\usepackage{mathtools}
\usepackage{color} 
\usepackage{hyperref} 
\usepackage{textcomp} 
\usepackage{lmodern} 
\usepackage[T1]{fontenc}

\begin{document}

\title{Measured velocity spectra and neutron densities of the PF2 ultracold-neutron beam ports at the Institut Laue--Langevin}

\author{Stefan D\"{o}ge}
\email[Corresponding author Stefan Doege: ]{stefan.doege@tum.de}

\affiliation{Institut Laue--Langevin, 71 avenue des Martyrs, F-38042 Grenoble Cedex 9, France}
\affiliation{Physik-Department, Technische Universit\"{a}t M\"{u}nchen, D-85748 Garching, Germany}

\author{J\"{u}rgen Hingerl}
\affiliation{Institut Laue--Langevin, 71 avenue des Martyrs, F-38042 Grenoble Cedex 9, France}
\affiliation{Physik-Department, Technische Universit\"{a}t M\"{u}nchen, D-85748 Garching, Germany}

\author{Christoph Morkel}
\affiliation{Physik-Department, Technische Universit\"{a}t M\"{u}nchen, D-85748 Garching, Germany}


\begin{abstract}
Ultracold neutrons (UCNs) are a useful tool for fundamental physics experiments. They can be used to probe the lifetime of free neutrons, search for new CP violating processes and exotic interactions beyond the Standard Model, perform Ramsey spectroscopy, and carry out neutron-optical interference experiments. All of these experiments require high neutron count rates for good statistics. For optimal exploitation of experimental beam time, these experiments need to be prepared and, at times, even simulated in advance. To this end, it is crucial to know the velocity-dependent UCN flux at each beam position. Knowing the absolute neutron flux also allows for an absolute calibration of previously gathered data. Using the same time-of-fight experimental setup, we have measured the differential neutron flux of three out of the four UCN beam ports at the PF2 instrument at Institut Laue--Langevin, Grenoble. These beam ports are commonly used for UCN flux experiments and proof-of-principle tests.
\\
\\
Published online on 11 November 2019: \url{https://doi.org/10.1016/j.nima.2019.163112} \\ S. D\"{o}ge, J. Hingerl, C. Morkel, Nuclear Instruments and Methods in Physics Research, A 953 (2020) 163112 \\ \textcopyright\, 2019 Elsevier B.V. This manuscript version is made available under the \href{https://arxiv.org/licenses/nonexclusive-distrib/1.0/license.html}{arXiv non-exclusive distribution license}.
\end{abstract}


\maketitle

\section{Introduction}

Many fundamental physics experiments use ultracold neutrons (UCNs). These are neutrons with a kinetic energy low enough to be confined in material bottles or magnetic traps, typically $\lesssim 300$~neV ($v < 7.6$~m/s)~\cite{ignatovich:1990,golub:1991}. This property allows for long observation times and makes UCNs particularly interesting for three types of experiments: storage experiments determining the lifetime of the free neutron, the search for a non-zero electric dipole moment of the neutron (nEDM), and constraining dark matter candidates~\cite{dubbers:2011}. UCNs are also frequently used in transmission experiments~\cite{doege:2015} as well as low-energy experiments that probe gravitation on the micrometer scale~\cite{nesvizhevsky:2002,cronenberg:2018} and neutron-optical phenomena~\cite{frank:2016}.

Various UCN converters are operational throughout the world. The oldest such converter, and by far the most often used for fundamental physics research until the present, is the UCN ``Turbine'' (instrument PF2) at the Institut Laue--Langevin~\cite{steyerl:1975,steyerl:1986} with its four UCN beam ports. It slows very cold neutrons (VCN) down to UCN energies by reflecting them off receding polished metal blades exploiting the Doppler effect. The design of the Turbine favors experiments requiring a high UCN flux over those requiring a high UCN density.

\section{Previous Measurements}

Steyerl et al. carried out UCN flux (also called current density) and density measurements after the Turbine had been installed \cite{steyerl:1986,steyerl:1989-ucn-sources}. In these experiments, both flux and density \emph{inside} the Turbine vessel were determined. However, no systematic measurement has yet been taken of the UCN \emph{flux} available \emph{outside} the vessel at the four PF2 beam ports (MAM, UCN, EDM, TEST), nor have these beam ports been compared with one another.

In 1999, the group led by A.~V.~Strelkov from JINR Dubna (Russia) measured the neutron \emph{density} of the EDM, UCN, and MAM beams using a spherical copper storage vessel with a volume of 27~liters ($v_\text{crit}^\text{Cu}=5.7$~m/s) and a pinhole exit~\cite{strelkov:2019}. Unfortunately, these data were never published.

Recently, a standardized steel storage vessel with a neutron-optical potential of 188~neV ($v_\text{crit}^\text{steel}=6.0$~m/s) on the inside walls was used to measure the UCN \emph{density} at one neutron beam port of the Turbine (PF2-EDM). The neutron densities at the exits of other UCN converters, based on solid deuterium and liquid helium as conversion media, were measured using the same storage vessel and then compared with one another~\cite{ries:2017}.

Many experiments, especially long-term experiments, will benefit from the data of the velocity-dependent UCN fluxes at the Turbine's beam ports~\cite{doege:2016-3-14-370}. This data will facilitate the simulation and preparation of experiments before the setup is installed at the Turbine for its allocated beam time. The data will also serve as a retroactive calibration standard for past experiments and as a reference for future instrument upgrades.

\section{The Experimental Setup}

\begin{figure}[!h]
\includegraphics[width=1.00\columnwidth]{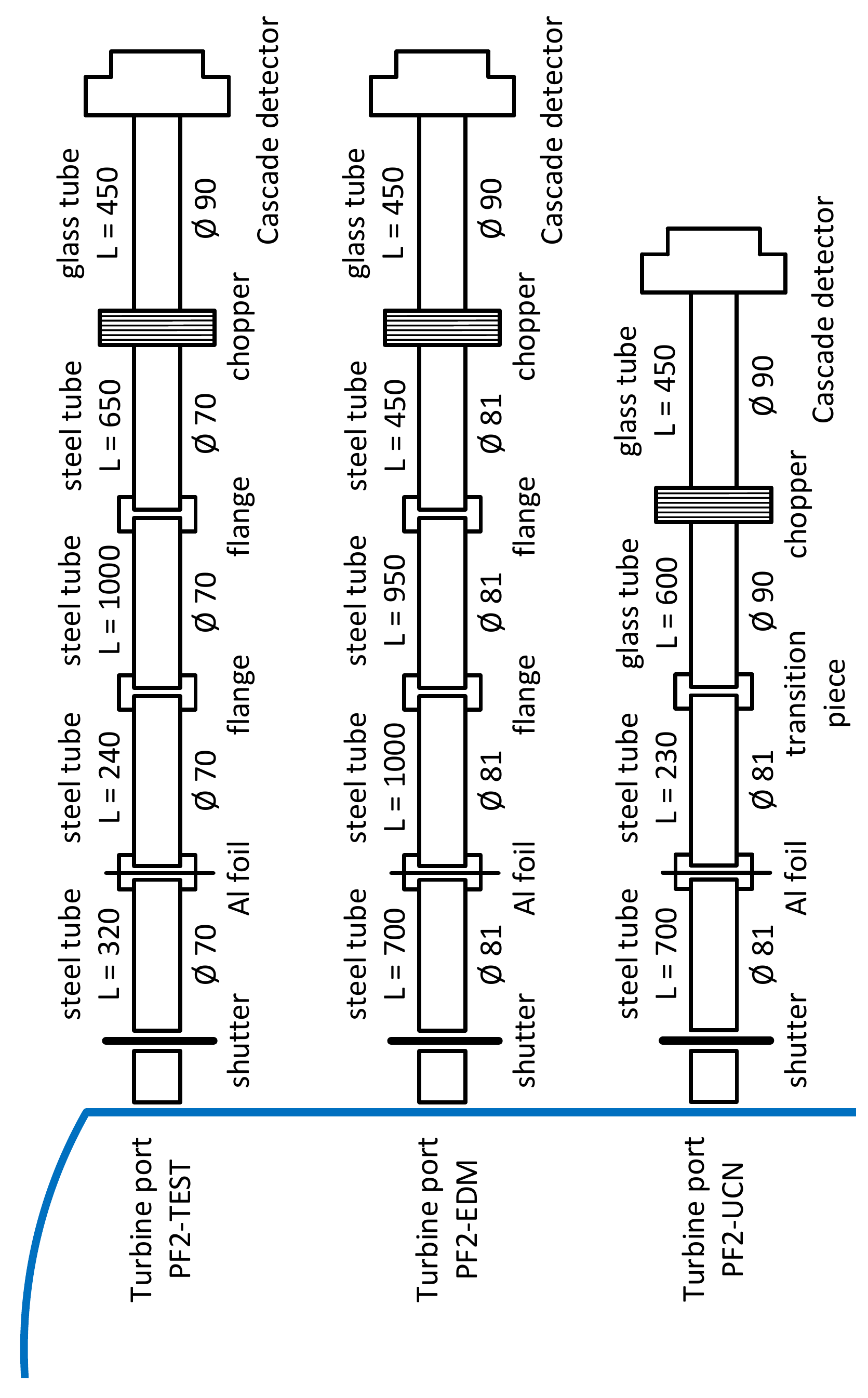}
\caption[Turbine beam tube configurations]{Beam tube configurations used for comparing the flux of the three Turbine beam ports PF2-TEST, PF2-EDM, and PF2-UCN with one another. The Turbine's safety foil is indicated as ``Al foil''. For each beam tube, the length $L$ and outer diameter $\varnothing$ are given in millimeters.}\label{fig:turbine-configurations}
\end{figure}

For the comparative measurement of neutron spectra, the UCN ports of the Turbine were equipped with beam tube configurations that are often used by experimenters, see Fig.~\ref{fig:turbine-configurations}. The beam port PF2-MAM was permanently occupied by the long-term experiment ``Gravitrap''~\cite{serebrov:2018} and thus its spectrum could not be measured.

For safety reasons, the vacuum in the Turbine's beam guides is separated from the Turbine vacuum by a 100~$\mu$m thick AlMg3 foil. The neutron guides between the safety foil of the Turbine and the chopper were standard NiMo-coated electropolished stainless steel tubes of various lengths and with outer diameters as indicated in Fig.~\ref{fig:turbine-configurations}. The tubes all had a wall thickness of 2~mm. Their transmissivity was measured to be 95\% per meter for the UCN spectrum of the Turbine.

Connected to the last steel guide tube was an UCN chopper similar to the one developed by Lauer~\cite{lauer:2010} but with 1~mm thick titanium grids, a NiMo-coated glass tube of 80~mm inner and 90~mm outer diameter, and a Cascade neutron detector~\cite{klein:2011}, which together constituted the time-of-flight (TOF) geometry of this experiment. The transmissivity of the glass tubes for neutrons in the UCN range was 87\% per meter, and the neutron flight path had a length of $d_\text{TOF}=458$~mm. As the neutron beam was uncollimated, the TOF method measured the neutrons' velocity component along the beam axis, $v_\text{z}$. All spectra were measured at the height of the Turbine exits and without any bends in the neutron guides.

The absolute efficiency of the Cascade detector was determined to be 34$\pm$5\%~\cite{hingerl:2019,doege:2019-phd} across the Turbine's neutron spectrum. This value, together with the beam tubes' transport efficiencies, and the chopper's duty cycle of 3.94\% and geometric opening of 36\% allowed the measured UCN transmission to be extrapolated back to the position of the aluminum safety foil and the absolute differential neutron flux $\phi(v)$ at the Turbine's beam ports to be obtained. All in all, the beamline and TOF setup reduce the usable number of UCNs to about 0.5\% of the original flux at the position of the safety foil.

All measurements were carried out at a thermal reactor power of 55.8$\pm$0.2~MW and with the Turbine operating in time-sharing mode. This means that after each measurement run of 120~s, the neutron switch could be requested by users on another beamline. To ensure equal background conditions at each measurement run, the measurement was only started after the beam guide had been filled with neutrons for 12~s.

\begin{figure}[!h]
\includegraphics[width=1.00\columnwidth]{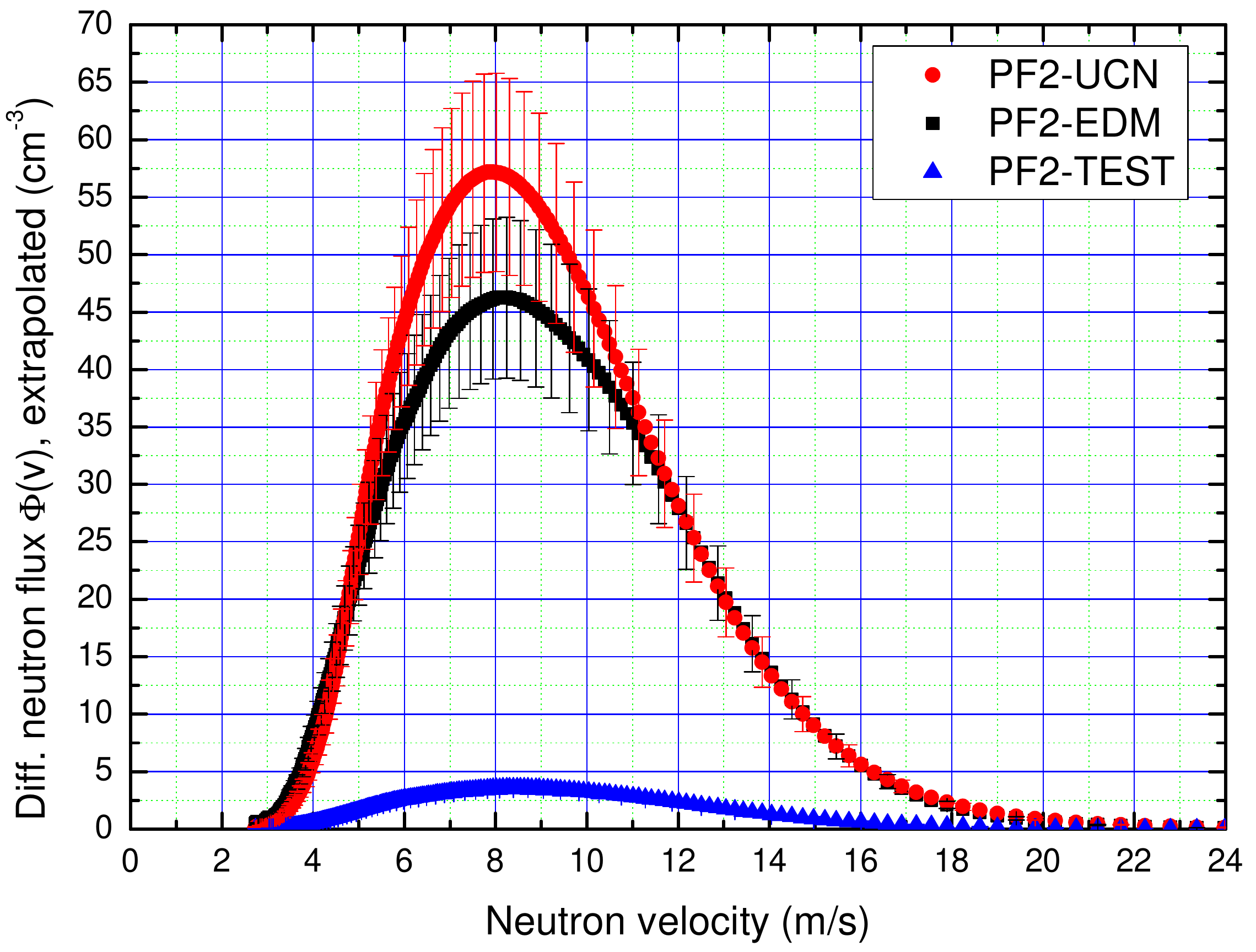}
\caption[Turbine beam ports compared]{Comparison of the Turbine's differential neutron fluxes $\phi (v)$ at the beam ports PF2-{UCN}, PF2-{EDM}, and PF2-{TEST} at the position of the respective safety foil. Of the error bars, a systematic error of 15\% is due to the uncertainty of the absolute Cascade detector efficiency for UCNs. The peak velocities can be calculated using Eq.~\ref{eq:peak-velocity} and the respective spectral temperature from Tab.~\ref{tab:densities}. Some error bars have been removed for better legibility.}\label{fig:turbine-ports-compared}
\end{figure}

\begin{widetext}

\begin{table}
\caption{\label{tab:densities} Neutron densities $N$, fluxes $\Phi$, and temperatures $T$ extracted from the Maxwell--Boltzmann (MB) fits and the real experimental spectra of the beam ports PF2-{UCN}, PF2-{EDM}, and PF2-{TEST}. The uncertainty indicated for the data in column 1 is the fit error. The fit error for the temperature in column 2 is of the order of $10^{-2}$~mK. All experimental data have a relative systematic uncertainty of $\pm 15$\%, which stems from the uncertainty of the neutron detector's detection efficiency. Statistical errors are negligible compared to the systematic errors. The critical velocities are $v_\text{crit}^\text{steel}=6.0$~m/s and $v_\text{crit}^\text{Ni-58}=8.1$~m/s.}
\begin{tabular}{|l||l|l|l||l|l|l|l|}\hline
  Name of & Tot. UCN density & Spectral & Integral neutron & Integral neutron & Tot. UCN density & $N\left(v_\text{crit}^\text{steel}\right)$ & $N\left(v_\text{crit}^\text{Ni-58}\right)$ \\
	beam port & $N_0$ [cm$^{-3}$] & temp. $T$ [mK] & flux $\Phi$ [cm$^{-2}$s$^{-1}$] & flux $\Phi$ [cm$^{-2}$s$^{-1}$] & $N_0$ [cm$^{-3}$] & [cm$^{-3}$] & [cm$^{-3}$] \\\hline
 & \multicolumn{3}{|c||}{Maxwell--Boltzmann fit} & \multicolumn{4}{|c|}{Experimental} \\\hline\hline
	PF2-UCN & $61.0\pm 0.1$ & 2.6 & 44,800 & 41,135 & 49.3 & 10.4 & 26.1 \\\hline
	PF2-EDM & $50.1\pm 0.1 $ & 2.8 & 38,400 & 36,307 & 43.3 & 9.87 & 22.4 \\\hline
	PF2-TEST & $3.86\pm 0.01$ & 3.0 & 3,060 & 2,928 & 3.37 & 0.69 & 1.62 \\\hline
\end{tabular}
\end{table}

\end{widetext}

\section{Experimental Results and Calculations}

The Turbine ports' neutron spectra, which extend beyond the UCN range into the VCN energy range, are shown in Fig.~\ref{fig:turbine-ports-compared}. They were corrected for detector efficiency and chopper duty cycle, and extrapolated back to the position of the Turbine's safety foil. 

The measured differential neutron flux $\phi(v)$ (in units of cm$^{-3}$) from Fig.~\ref{fig:turbine-ports-compared} can be well approximated by a Maxwell--Boltzmann (MB) distribution $f^\text{MB}(v)$ for the neutron density $N(v)=N_0\times f^\text{MB}(v)$, where $\phi(v) = N(v)\times v$, and thus
\begin{equation}\label{eq:maxwell-boltzmann}
\phi(v)=N_0 4\pi \left( \frac{m_\text{n}}{2\pi k T} \right)^{3/2} v^3 \times \text{exp}\left( \frac{m_\text{n} v^2}{2kT} \right).
\end{equation}

\noindent
In Eq.~\ref{eq:maxwell-boltzmann}, $N_0$ represents the total neutron density (in units of cm$^{-3}$) over the entire spectrum, $m_n$ the neutron's mass, $k$ the Boltzmann constant, $T$ the temperature of the Maxwell--Boltzmann shaped neutron spectrum, and $v$ the neutron velocity. $f^\text{MB}(v)$ is normalized to unity, in other words,
\begin{equation}\label{eq:MB-normal}
\int_{0}^{\infty} N(v) \text{d}v = N_0.
\end{equation}

By integrating Eq.~\ref{eq:MB-normal} not over the entire spectrum but rather over an arbitrary velocity range, e.g. from 0~m/s to $v_\text{crit}$, the neutron density $N(v_\text{crit})$ over that range can be calculated. Here, 
\begin{equation}
v_\text{crit}=\sqrt{\frac{2 U_\text{opt}}{m_\text{n}}},
\end{equation}

\noindent
where $U_\text{opt}$ is the neutron-optical potential of a given material~\cite{ignatovich:1990,golub:1991}. When calculating the neutron density inside an UCN storage bottle, $U_\text{opt}$ is determined by the storage bottle's inner wall coating, to which the neutrons are exposed. Materials of choice are often steel, copper, nickel, diamond-like-carbon (DLC), or fomblin grease.

The data from Fig.~\ref{fig:turbine-ports-compared} have been fitted with the function $\phi (v) = N(v)\times v$ , see Eq.~\ref{eq:maxwell-boltzmann}, over the velocity interval $(6~\text{m/s},\infty)$. From this fit, the total neutron densities $N_0$, temperatures $T$, and integral neutron fluxes $\Phi$ shown in Tab.~\ref{tab:densities} could be extracted. It must be noted, however, that the experimental data for $v< 6~\text{m/s}$ lie lower than the MB-based fit and the fit, therefore, overestimates to some extent both the integral neutron flux and the total neutron density of the spectrum. This is due to the critical velocity of both the aluminum safety foil and the aluminum entrance window of the Cascade detector ($v_\text{crit}^\text{Al}=3.2~\text{m/s}$), which prevent the slowest neutrons from exiting the Turbine into the beam guide and entering the detector. However, the rest of the spectrum agrees remarkably well with the MB fit, as demonstrated by the minuscule fit error of the total neutron density shown in Tab.~\ref{tab:densities}.

The peak of the differential neutron flux is found at $\phi(v^{\ast})$ in Fig.~\ref{fig:turbine-ports-compared}, while the most probable velocity of the Maxwell-Boltzmann spectrum is found at $f^\text{MB}(\hat{v})$. Both are related to each other by
\begin{equation}\label{eq:velocities}
v^{\ast} = \sqrt{\frac{3}{2}}\hat{v},
\end{equation}

\noindent
which can easily be demonstrated by calculating the first derivatives $\partial / \partial v$ of both distributions.

Using the relation
\begin{equation}\label{eq:peak-velocity}
\hat{v}=\sqrt{\frac{2kT}{m_\text{n}}},
\end{equation}

\noindent
one can convert the temperature of the neutron spectra, see Tab.~\ref{tab:densities}, into each spectrum's peak velocity $\hat{v}$, and also rewrite Eq.~\ref{eq:maxwell-boltzmann} as
\begin{equation}\label{eq:simple-mb}
\phi(v)=N_0 \frac{4}{\sqrt{\pi}} \times \left( \frac{v}{\hat{v}} \right)^{3} \times \text{exp}\left( -\left[ \frac{v}{\hat{v}}\right]^2 \right).
\end{equation}

Calculated for the neutron flux peak $\phi (v^{\ast})$ using the relation from Eq.~\ref{eq:velocities}, Eq.~\ref{eq:simple-mb} simplifies to
\begin{equation}
\phi (v^{\ast}) = N_0 \frac{4}{\sqrt{\pi}} (3/2)^{3/2} \times \text{exp}(-3/2) = N_0 \times 0.9251.
\end{equation}

The total neutron density $N_0$ of a MB distributed neutron spectrum is thus related to the peak of the differential flux by a simple scaling factor. 

The integral neutron flux $\Phi$ (in units of cm$^{-2}\times$s$^{-1}$) of the spectrum up to a critical velocity $v_\text{crit}$ can be calculated by integrating Eq.~\ref{eq:simple-mb}
\begin{equation}\label{eq:neutron-density}
\Phi = \int_{0}^{v_\text{crit}} N(v) \times v \text{d}v = N_0 \underbrace{ \frac{\int_{0}^{v_\text{crit}} f^\text{MB}(v)\times v \text{d}v}{\int_{0}^{v_\text{crit}} f^\text{MB}(v) \text{d}v} }_{\langle v \rangle},
\end{equation}

\noindent
with the analytical solution of the integral in the numerator equal to
\begin{equation}
\hat{v}\times \frac{2}{\sqrt{\pi}}\times \left( 1 - \left[ 1 + \left( \frac{v_\text{crit}}{\hat{v}} \right)^2 \right]\times \text{exp}\left[ -(v_\text{crit}/\hat{v})^2\right] \right),
\end{equation}

\noindent
and the solution of the integral in the denominator being
\begin{equation}
\text{erf}\left( \frac{v_\text{crit}}{\hat{v}} \right) - \frac{2}{\sqrt{\pi}}\times \left( \frac{v_\text{crit}}{\hat{v}} \right) \times \text{exp}\left( -[v_\text{crit}/\hat{v}]^2\right).
\end{equation}

Eq.~\ref{eq:neutron-density} is tantamount to calculating the mean velocity $\langle v \rangle$ of the velocity distribution and multiplying it by the total neutron density. This is exactly the definition of the integral neutron flux $\Phi$.

For a perfect MB distribution, the weighted mean velocities $\langle v \rangle$ from the aluminum cut-off (3.2~m/s) to the steel cutoff (6.0~m/s) are 4.74~m/s for the PF2-UCN beam, 4.75~m/s (PF2-EDM), and 4.76~m/s (PF2-TEST). The weighted mean velocities extracted from the experimental spectrum are 5.13~m/s for the PF2-UCN beam, 5.01~m/s (PF2-EDM), and 5.04~m/s (PF2-TEST). They lie slightly higher than the velocities from the MB fit due to the aluminum cut-off of the slowest UCNs. At the aluminum safety foil, even UCNs slightly faster than the critical velocity of aluminum are reflected if they impinge on the foil at an angle other than 90\textdegree. The roughness of the aluminum foil also has an effect on the transmission of UCNs~\cite{doege:2019-phd, doege:2020-foils}.

Tab.~\ref{tab:densities} shows the values for neutron densities (total and up to the critical velocities of steel and Ni-58), fluxes, and temperatures of the measured spectra of the beam ports PF2-UCN, PF2-EDM, and PF2-TEST as extracted by
\begin{enumerate}
	\item Maxwell--Boltzmann (MB) fits using the mathematical relations explained above
  \item numerical integration and direct extraction from the experimental data.
\end{enumerate}

It is easy to see that both the beam ports PF2-UCN and PF2-EDM provide comparable UCN densities. Both are thus equally suitable for UCN storage experiments.

The UCN density up to the critical velocity of steel $v_\text{crit}^\text{steel}$ at the EDM beam port (\emph{outside} the Turbine), $9.87\pm 1.48$~cm$^{-3}$, is about a factor of two larger than the density of $4.61\pm 0.02$~cm$^{-3}$ measured in a storage experiment for a similar beamline configuration, including the aluminum safety foil, as reported by Ries et al.~\cite{ries:2017}. The difference might be due to differences in the UCN transport and detection efficiencies between these two types of experiments.

Steyerl et al.~\cite{steyerl:1986} reported the UCN density up to 6.2~m/s \emph{inside} the Turbine to be 87~cm$^{-3}$ as derived from TOF data and to be 36~cm$^{-3}$ as measured with an iron storage bottle. Here, again, the neutron density measured by the TOF method lies about a factor of two higher than the density measured with a storage bottle.

\section{Conclusions}

Using the same experimental equipment, we have measured and compared the differential neutron fluxes at three out of the four beam ports of the UCN Turbine at the Institut Laue--Langevin. From these data, we were able to calculate the integral neutron fluxes and extract the total neutron densities up to any arbitrary critical neutron velocity $v_\text{crit}$. Our measured values for the PF2-UCN and PF2-EDM beams indicate that both have a similar UCN density below the steel cut-off (6.0~m/s) of 10.4~cm$^{-3}$ and 9.87~cm$^{-3}$, respectively. These are in line with earlier measurements of the UCN density inside the Turbine vessel. The density of UCNs below the steel cut-off at the PF2-TEST beam is 0.69~cm$^{-3}$. These are the first published comparative neutron density measurements outside the Turbine.

The experimental results for this paper were produced as part of the Ph.D. theses of Stefan D\"oge~\cite{doege:2019-phd} and Tobias Rechberger~\cite{rechberger:2018}.

\begin{acknowledgments}
We wish to thank the instrument scientists of PF2 for their advice and Dr.~Tobias Rechberger for support during the experiment. The PhD thesis of S.~D. was done within a collaboration between the Institut Laue--Langevin (ILL), Grenoble, France and Technische Universit\"{a}t M\"{u}nchen, Munich, Germany. It received financial support from both the ILL and FRM~II/ Heinz Maier-Leibnitz Zentrum (MLZ), Garching, Germany. Furthermore, J.~H. and S.~D. acknowledge funding from Dr.-Ing. Leonhard-Lorenz-Stiftung, Munich, under grant no. 930/16.
\end{acknowledgments}


%

\end{document}